\documentclass[aps,prl,reprint,superscriptaddress]{revtex4-1}
\usepackage{CJK}

\usepackage{upgreek}
\usepackage{amsmath}
\usepackage{amssymb}
\usepackage{graphicx}
\usepackage{natbib}
\usepackage{amsmath}
\usepackage{epigraph}
\usepackage{svg}
\begin{document}
\begin{CJK*}{GB}{} 

\title{Detection of Low-Energy Electrons with Transition-Edge Sensors}

\author{Carlo Pepe}
\affiliation{Istituto Nazionale di Ricerca Metrologica, Strada delle Cacce 91, 10135 Torino, Italy}
\affiliation{Politecnico di Torino - Dipartimento di Elettronica e Telecomunicazioni, Corso Duca degli Abruzzi 24, 10129 Torino, Italy}

\author{Benedetta Corcione}
\affiliation{Sapienza Universit\`{a} di Roma, Piazzale Aldo Moro 2, 00185 Rome, Italy}
\affiliation{Istituto Nazionale di Fisica Nucleare - Sezione di Roma, Piazzale Aldo Moro 2, 00185 Rome, Italy}

\author{Francesco Pandolfi}
\email{francesco.pandolfi@roma1.infn.it}
\affiliation{Istituto Nazionale di Fisica Nucleare - Sezione di Roma, Piazzale Aldo Moro 2, 00185 Rome, Italy}

\author{Hobey Garrone}
\affiliation{Istituto Nazionale di Ricerca Metrologica, Strada delle Cacce 91, 10135 Torino, Italy}
\affiliation{Politecnico di Torino - Dipartimento di Elettronica e Telecomunicazioni, Corso Duca degli Abruzzi 24, 10129 Torino, Italy}

\author{Eugenio Monticone}
\affiliation{Istituto Nazionale di Ricerca Metrologica, Strada delle Cacce 91, 10135 Torino, Italy}

\author{Ilaria Rago}
\affiliation{Istituto Nazionale di Fisica Nucleare - Sezione di Roma, Piazzale Aldo Moro 2, 00185 Rome, Italy}

\author{Gianluca Cavoto}
\affiliation{Sapienza Universit\`{a} di Roma, Piazzale Aldo Moro 2, 00185 Rome, Italy}
\affiliation{Istituto Nazionale di Fisica Nucleare - Sezione di Roma, Piazzale Aldo Moro 2, 00185 Rome, Italy}

\author{Alice Apponi}
\affiliation{Dipartimento di Scienze Universit\`a degli Studi Roma Tre, and Istituto Nazionale di Fisica Nucleare - Sezione di Roma Tre, Via della Vasca Navale 84, 00146 Rome, Italy}

\author{Alessandro Ruocco}
\affiliation{Dipartimento di Scienze Universit\`a degli Studi Roma Tre, and Istituto Nazionale di Fisica Nucleare - Sezione di Roma Tre, Via della Vasca Navale 84, 00146 Rome, Italy}

\author{Federico Malnati} 
\affiliation{Dipartimento di Fisica, Universit\`a di Torino, via Pietro Giuria 1, 10125, Torino, Italy}

\author{Danilo Serazio}
\affiliation{Istituto Nazionale di Ricerca Metrologica, Strada delle Cacce 91, 10135 Torino, Italy}

\author{Mauro Rajteri}
\affiliation{Istituto Nazionale di Ricerca Metrologica, Strada delle Cacce 91, 10135 Torino, Italy}

\begin{abstract} 

We present the first detection of electrons with kinetic energy in the 100~eV range with transition-edge sensors (TESs).  This has been achieved with a $(100\times 100)$~$\upmu$m$^2$ Ti-Au bilayer TES, with a critical temperature of about 84~mK. The electrons are produced directly in the cryostat by an innovative cold  source based on field emission from vertically-aligned multiwall carbon nanotubes. We obtain a Gaussian energy resolution between 0.8 and 1.8~eV for fully-absorbed electrons in the $(90-101)$~eV energy range, which is found to be compatible with the resolution of this same device for photons in the same energy range. This work opens new possibilities for high-precision energy measurements of low-energy electrons.


\end{abstract}

\maketitle


Transition-edge sensors (TESs) are highly sensitive micro-calorimeters capable of high-resolution single-photon counting across a wide energy spectrum \cite{irwin1995application,cunningham2002high}. The detection scheme is based on the absorption of the incoming photons in a thin superconducting film, in which their energy is transformed into heat. By operating a TES device at its critical temperature $T_\text{C}$, even small variations in temperature lead to measurable changes in its electrical resistance, owing to the steep transition between the superconducting regime and the normal-conduction one. TES devices have been capable of achieving single-photon Gaussian energy resolutions below 50~meV for 0.8~eV photons \cite{hattori2022optical, lolli}. 

In principle, this detection scheme should also be sensitive to low-energy electrons absorbed in the superconducting film, as seems to be confirmed by simulations~\cite{patel2021simulation}. However, there is currently very limited research on the use of TES devices for electron detection, except for a recent result~\cite{patel2024} on electrons in the 300-2000~eV energy range, which achieves a Gaussian energy resolution $\sigma_{\mathrm{e}} > 17$~eV. Low-energy electrons have been detected with other detection schemes, such as micro-channel plates~\cite{Apponi:2022nxn}, which have only very poor energy resolution, and also feature geometrical inefficiencies due to their non-unitary fill factor; and silicon detectors, such as avalanche photo-diodes~\cite{Apponi:2020upy} and silicon drift detectors~\cite{GUGIATTI2020164474}, which are characterized by the presence of a dead layer at the detector entrance, in which low-energy electrons are absorbed before generating a signal. On the other hand TES devices offer, in principle, a detection scheme without dead layers and with unitary fill factor, therefore with the potential for high efficiency and excellent intrinsic energy resolution on low-energy electrons. This would be of great interest for a wide range of experiments, such as, for example, the PTOLEMY project \cite{Betti:2018bjv, Betti_2019, Apponi:2021hdu}, which plans to search for the cosmic neutrino background by analyzing the endpoint of the beta decay of tritium with unprecedented electron energy resolution: the target of the experiment is to achieve an electron energy resolution of 50~meV on 10~eV electrons.

\begin{figure}[tb]
\includegraphics[width=0.45\textwidth]{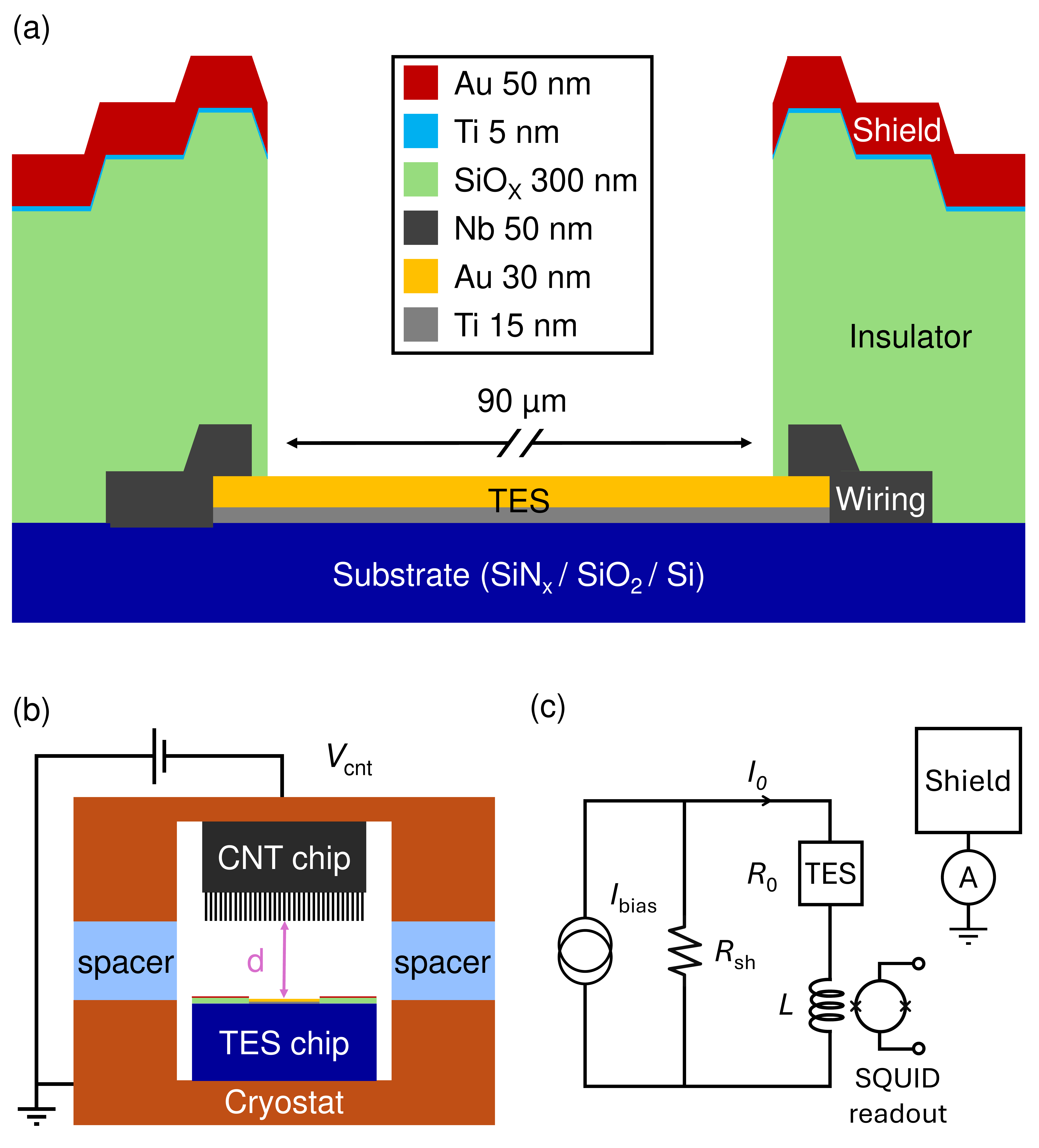}
\caption{TES layout and experimental setup. (\textbf{a}): Schematic view of the TES device and its shield layer (y-axis  to scale, x-axis not to scale). (\textbf{b}):  Schematic view of the setup used for electron counting: the carbon nanotubes (CNTs) are hosted on the top copper plate and oriented with their tips pointing towards the TES.  (\textbf{c}): Schematic view of the electrical circuit. The TES operates in electrothermal feedback mode. Current signals produced by single electrons are read using a SQUID array. The shield around the TES is connected to a picoammeter to measure the current of the electron from the CNT. }
\label{fig:setup}
\end{figure}

In this work, electrons are produced by field emission from vertically-aligned, multi-wall carbon nanotubes~(CNTs). This innovative cold-source solution overcomes the issues in interfacing standard hot-filament-based electron sources with the TES working at cryogenic temperature, thus allowing, for the first time, to place the source directly inside the cryostat, close to the TES. 

The results presented in the following were obtained in the Innovative Cryogenic Detectors Laboratory of Istituto Nazionale di Ricerca Metrologica (INRiM) in Torino (Italy) \cite{pepe2023superconducting}. The TES detectors operate inside an adiabatic demagnetization refrigerator cryostat, at a stable bath temperature $T_{\mathrm{bath}} = 41.7 \pm 0.2$~mK, as measured by a ruthenium oxide thermal sensor. Since electrons in motion are affected by magnetic fields, and TESs and DC-SQUIDs are also sensitive to magnetic fields, a cryogenic magnetic shield was installed within the cryostat around the working area to suppress any magnetic interference that could degrade the sensor performance and alter electron trajectories.

\begin{figure}[tb]
\includegraphics[width=0.4\textwidth]{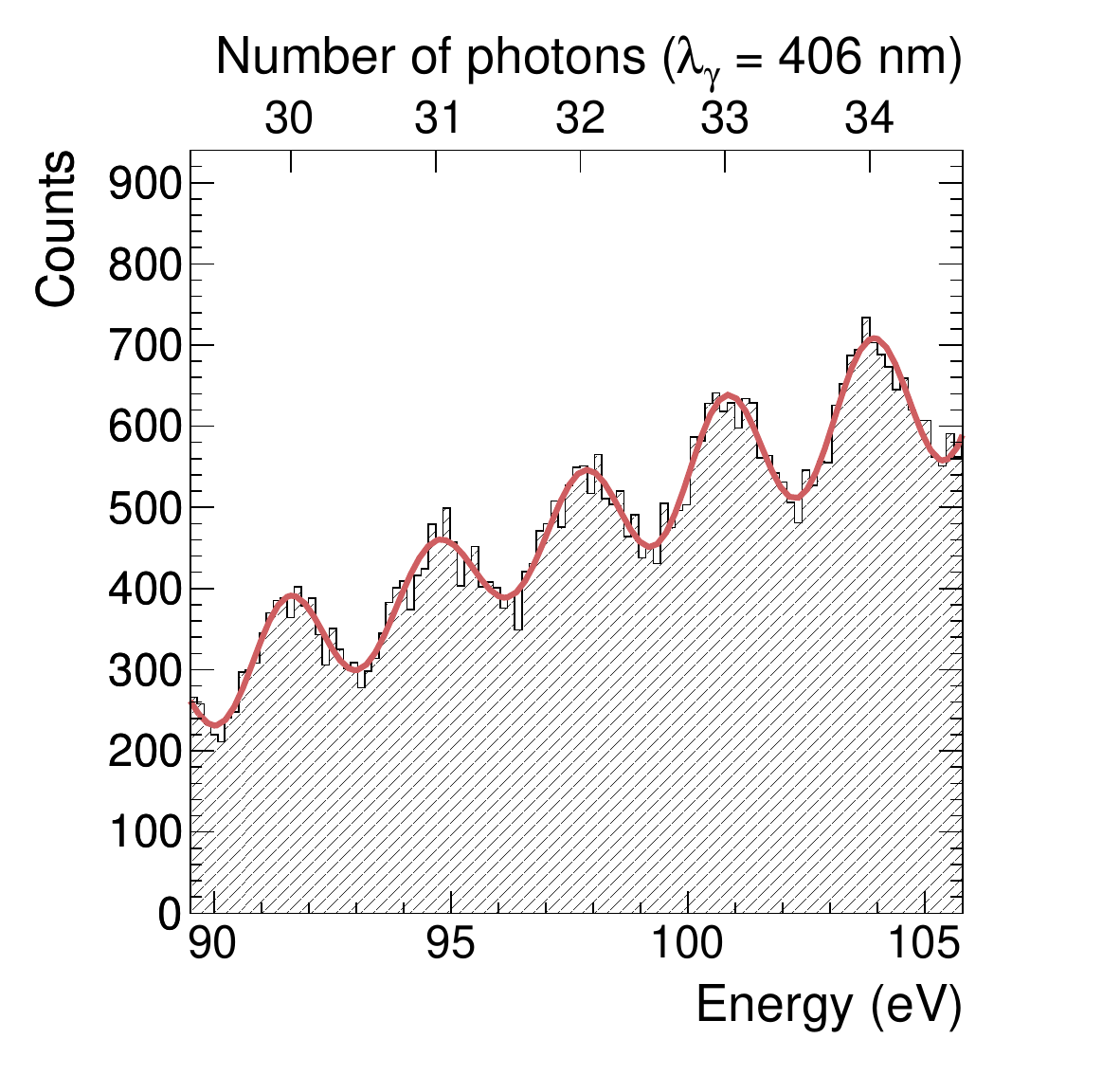}
\caption{Results of the TES optical characterization with $\lambda_{\gamma} = 406$~nm photons. The histogram of the pulse amplitudes report peaks corresponding to a number of photons $30 \leq N_{\gamma} \leq 34$ which are clearly distinguishable, and fitted with a sum of Gaussian functions (red line).}
\label{fig:photons}
\end{figure}

A schematic view of the layers which compose the TES and its surroundings is shown in panel (a) of Figure~\ref{fig:setup}. The TES device used in this work was fabricated at QR Lab, INRiM, has an area of $(100\times 100)$~$\mathrm{\upmu m}^2$,  and is a Ti-Au bilayer device, composed of a 15~nm layer of titanium covered by a 30~nm layer of gold, deposited by thermal evaporation on a silicon substrate covered by 150~nm of SiO$_2$ and 500~nm of SiN$_{\mathrm{x}}$ on both sides~\cite{monticone2021ti}. The wiring is done with 50~nm superconducting niobium strips, which were deposited on the substrate via sputtering. The TES has a critical temperature $T_\mathrm{C} = 84$~mK and was characterized in the $(3-143)$~eV energy range with 406~nm photons from a pulsed laser. Photons were directed onto the TES active area using a 9~$\upmu$m core silicon fiber aligned in free space. By varying the laser intensity, the number of photons per bunch ($N_{\gamma}$) could be adjusted within the range $0 < N_{\gamma} < 47$. The result of this characterization, in the energy range of interest for this work, is shown in Figure~\ref{fig:photons}: the peaks corresponding to $30 \leq N_{\gamma} \leq 34$ are clearly distinguishable, and fitted with a sum of Gaussian functions. The dark count of the TES was found to be negligible compared to the signal rate with the trigger threshold used in this work, as expected from this type of devices~\cite{manenti2024study}.

This TES design was adapted for electron detection by adding a shield layer, which is needed as the electron source has a significantly larger area: approximately ($3\times 3$)~mm$^2$ compared to the TES active area ($100\times 100$)~$\upmu$m$^2$. 
The shield leaves the TES active area exposed, but covers the area surrounding it, specifically the wiring and the substrate, as direct electron hits on the wiring would induce electrical noise, while on the insulating substrate would lead to charge build-up. 
The shield layer was produced by thermal evaporation, depositing an insulating layer consisting of 300~nm of amorphous silicon oxide (SiO$_\mathrm{x}$)~\cite{doi:10.1080/13642810008209760}, followed by a thin (5~nm) layer of titanium, and finally a 50~nm layer of gold. The thin titanium layer is necessary for best adhesion of gold to the SiO$_\mathrm{x}$. 



The electron source consists of a sample of vertically-aligned CNTs synthesized in the INFN laboratory `TITAN' at Sapienza University of Rome~\cite{nano13061081, YADAV2024169081, SARASINI2022110136, nano14010077}. The nanotubes were grown through chemical vapor deposition on a 500~$\upmu$m silicon substrate, and cover a surface of roughly $(3\times 3)$~mm$^2$. Thanks to the high geometrical field enhancement factor of their tips, nanotubes are capable of emitting electrons through quantum tunneling (field emission) without the necessity of applying very high voltages~\cite{fieldemission, LAHIRI20115411, semet2002field, lin2015field}. Furthermore, as field emission is quantic in nature, and not thermal, it does not generate heat and can therefore be used in a cryostat. A detailed characterization of field emission from CNTs at cryogenic temperatures will be the focus of a future work, currently in preparation.

The TES and the nanotubes were placed on two copper plates, facing each other, separated by 0.5~mm sapphire spacers, which ensure electrical insulation while guaranteeing a good degree of thermal conductance. The top copper plate, where the nanotubes are hosted, was connected to a power supply, and was provided negative voltage $V_{\mathrm{cnt}}$ in order to produce field-emission electrons; the bottom plate  was put in thermal contact with the cryostat, and grounded electrically through it. The distance between the tips of the nanotubes and the surface of the TES, in this setup, was $d = 612 \pm 32$~$\upmu$m. A schematic view of the setup is shown in panel (b) of Figure~\ref{fig:setup}.

In field emission the electrons tunnel through the potential barrier, and are therefore emitted at a potential $\phi_{\mathrm{cnt}}$ below the vacuum level, where $\phi_{\mathrm{cnt}}$ is the work function of the CNTs~\cite{PhysRevB.70.245410}. Furthermore, the difference between the work functions of the CNTs and the TES will create an effective field which will further correct the kinetic energy of the electrons. In formulas, the kinetic energy $E_{\mathrm{e}}$ of the electrons entering the TES is given by 
\begin{equation}
\label{eq:energy}
E_{\mathrm{e}} = eV_{\mathrm{cnt}} - \phi_{\mathrm{cnt}} + (\phi_{\mathrm{cnt}} - \phi_{\mathrm{tes}}) = eV_{\mathrm{cnt}} - \phi_{\mathrm{tes}}
\end{equation}
where $e$ is the elementary electric charge, and the TES work function was measured with ultraviolet photoemission spectroscopy in Roma Tre University's  LASEC lab, obtaining $\phi_{\mathrm{tes}} = 4.38 \pm 0.03$~eV. Moreover, we can assume that the source is monochromatic: the energy spread of the electrons, in fact, is expected to be of the order of \(kT\), which at cryogenic temperatures is only a few \(\upmu\)eV.

The electron emission from the CNTs was measured in two different ways: as a current $I_{\mathrm{cnt}}$ hitting the shield layer; and counting the signals due to single electrons hitting the TES. The electrical circuits of these two configurations are shown in panel (c) of Figure~\ref{fig:setup}. The $I_{\mathrm{cnt}}$ measurement was done by connecting a Keithley 6487 picoammeter to the shield layer, which served as a large-area metallic plate sensitive to the electron current hitting it. The signals generated by single electrons were measured by polarizing the TES on its transition using the current generator $I_\text{bias}$ and read out with a 16-SQUID series array~\cite{drung2007highly}, which is coupled to the inductance $L = 6$~nH. The TES working point was chosen to be $R_0 = 0.35 \cdot R_{\mathrm{N}}$, where $R_{\mathrm{N}} = 246$~m$\Omega$ is the resistance of the TES in its normal state. This working point was found to be optimal in terms of energy resolution during the optical characterization.
To ensure stable operation, the TES is operated in electrothermal feedback mode, which requires it to be voltage-biased \cite{irwin1995application}. This is achieved by connecting the TES in parallel to a shunt resistor $R_\text{sh} = 20$~m\(\Omega\),  significantly smaller than $R_0$.


\begin{figure}[tb]
\includegraphics[width=0.4\textwidth]{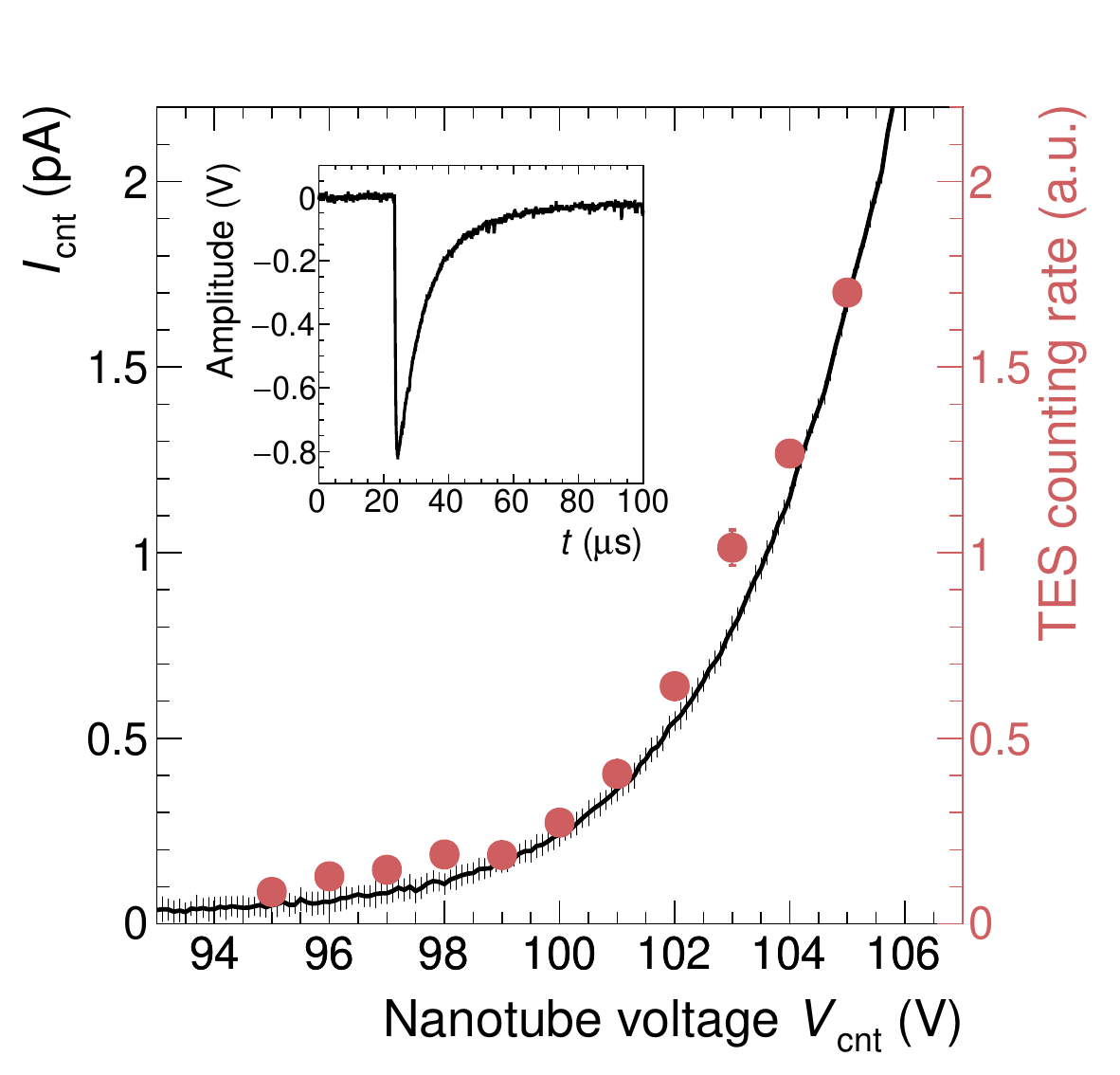}
\caption{Current $I_{\mathrm{cnt}}$ emitted by the nanotube source (black curve, left vertical scale), as a function of the negative voltage $V_{\mathrm{cnt}}$ provided to it, compared to the rate of pulses (red markers, right vertical scale) recorded by the TES. The inset shows a typical TES pulse-shape for \(V_{\text{cnt}} = 100\) V with a recovery time of approximately 10 \(\upmu\)s.
\label{fig:rate}}
\end{figure}

The $I_{\mathrm{cnt}}$ measurements, as a function of $V_{\mathrm{cnt}}$, are summarized by the black curve of Figure~\ref{fig:rate}. As can be seen, $I_{\mathrm{cnt}}$ exhibits an exponential rise, compatible with the Fowler-Nordheim theory on field emission~\cite{fowler1928electron}. Superimposed on the same plot with red markers is the rate of counts recorded by the TES in counting mode. As can be seen the increase in counting rate as a function of $V_{\mathrm{cnt}}$ follows the same exponential rise as  $I_{\mathrm{cnt}}$, therefore proving that the signals recorded by the TES are due to electrons. The inset in Figure~\ref{fig:rate} shows a typical TES signal for $V_{\mathrm{cnt}} = 100$~V, characterized by a rise time $\tau_+ \approx 200$~ns, and a recovery time $\tau_- \approx 10$~$\upmu$s.

\begin{figure}[tb]
\includegraphics[width=0.4\textwidth]{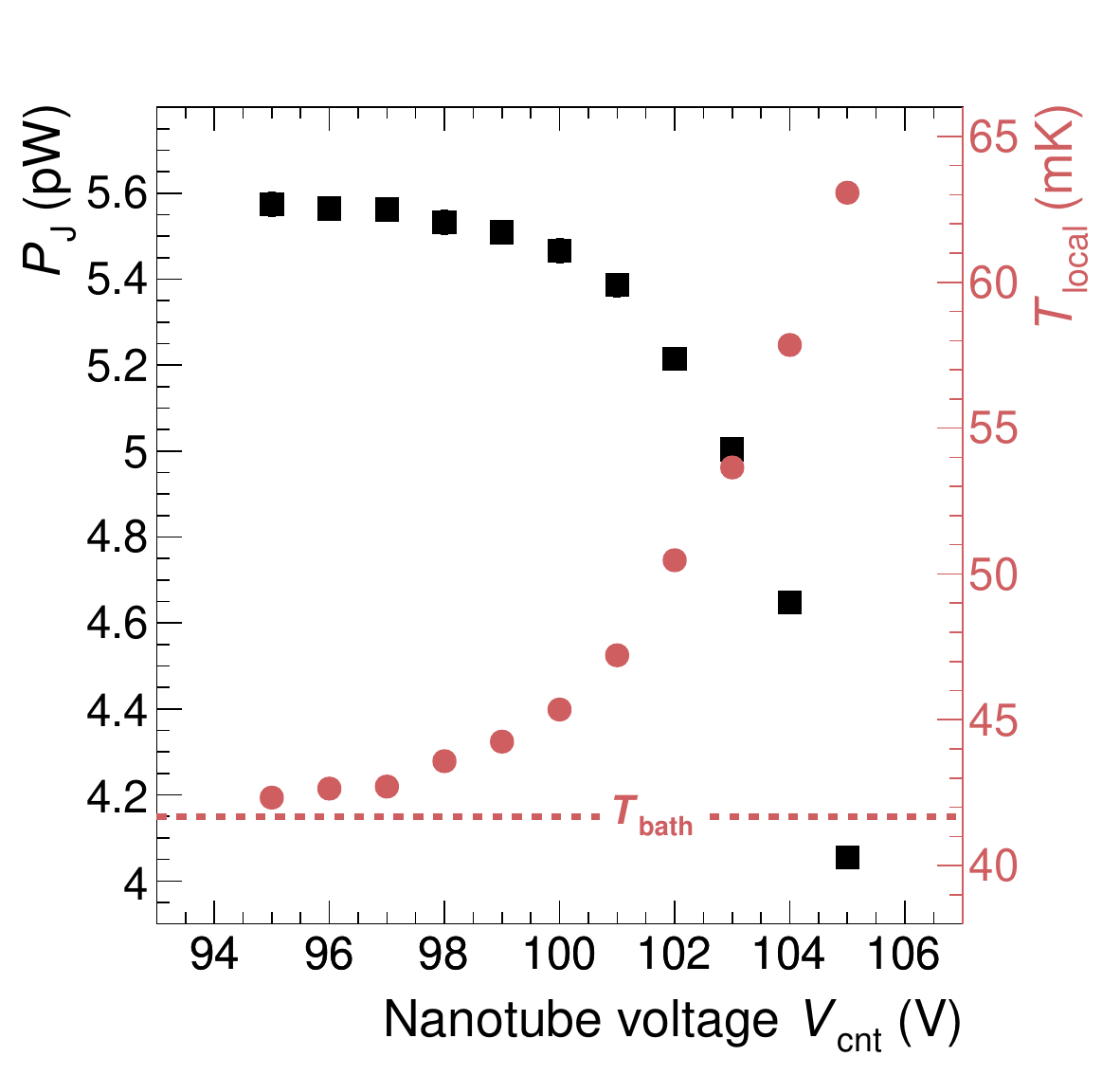}
\caption{Power $P_{\mathrm{J}}$ needed to bring the TES to its working point (black markers, left vertical scale) and local temperature $T_{\mathrm{local}}$ around the TES device (red markers, right vertical scale) for different values of \(V_{\mathrm{cnt}}\). The bath temperature $T_{\mathrm{bath}} = 41.7$~mK is indicated by the red dashed line.}
\label{fig:heat}
\end{figure}

A feature of Fowler-Nordheim emission is that the current of emitted electrons by the nanotubes depends on the electric field $|\vec{E}|$ in proximity of their tips, which in our planar configuration corresponds, in first approximation, to $|\vec{E}| = V_{\mathrm{cnt}}/d$.
At the same time, if we assume that electrons are emitted by the nanotubes with zero initial kinetic energy, \(V_{\mathrm{cnt}}\) also determines the kinetic energy of the electrons when entering the TES. Therefore, in this setup, the signal rate and electron energy are not independent parameters, as they both depend on \(V_{\mathrm{cnt}}\).

The electrons are emitted from a relatively large area compared to the TES, and the additional heating created by the electrons hitting its nearby environment must be considered. In general, the Joule power $P_{\mathrm{J}}$ required to bring the TES to its critical temperature \(T_\text{C}\) is given by~\cite{irwin2005transition}: 
\begin{equation}
\label{eq:power}
P_\text{J}(T_\text{local}) = I_\text{0}^2 R_{\mathrm{0}}(T_{\mathrm{local}}) = \kappa(T_\text{C}^n - T_\text{local}^n)
\end{equation}
where: $I_0$ is the current flowing in the TES; the exponential parameter and the coupling constant of this device were respectively measured to be \(n = 4.78 \pm 0.07\) and \(\kappa= (10 \pm 1) \cdot 10^{-7}\) W\(\text{K}^{-n}\); and $T_\text{local}$ is the local temperature in proximity of the TES. When the CNTs are off (\(V_{\mathrm{cnt}} = 0\)), or at low \(V_{\mathrm{cnt}}\), the electron rate and the energy deposited are minimal, therefore $T_\text{local} \approx T_\text{bath}$. As \(V_{\mathrm{cnt}}\) increases, the rate of electrons and the associated energy deposition in the vicinity of the TES increase, raising $T_\text{local}$.  This effect is shown in Figure~\ref{fig:heat}, where the black markers represent $P_{\mathrm{J}}$, as measured from $I_{\mathrm{0}}$ and $R_{\mathrm{0}}$, and the red markers represent $T_{\mathrm{local}}$, as obtained by inverting Equation~(\ref{eq:power}). As can be seen, when operating the electron source at \(V_{\mathrm{cnt}} = 105\) V, the local temperature is already $T_{\mathrm{local}} > 63$~mK, over 20~mK above $T_{\mathrm{bath}}$. This implies that results at different \(V_{\mathrm{cnt}}\) values are not rigorously comparable, as the TES operates under slightly different conditions.


\begin{figure}[bt]
\includegraphics[width=0.4\textwidth]{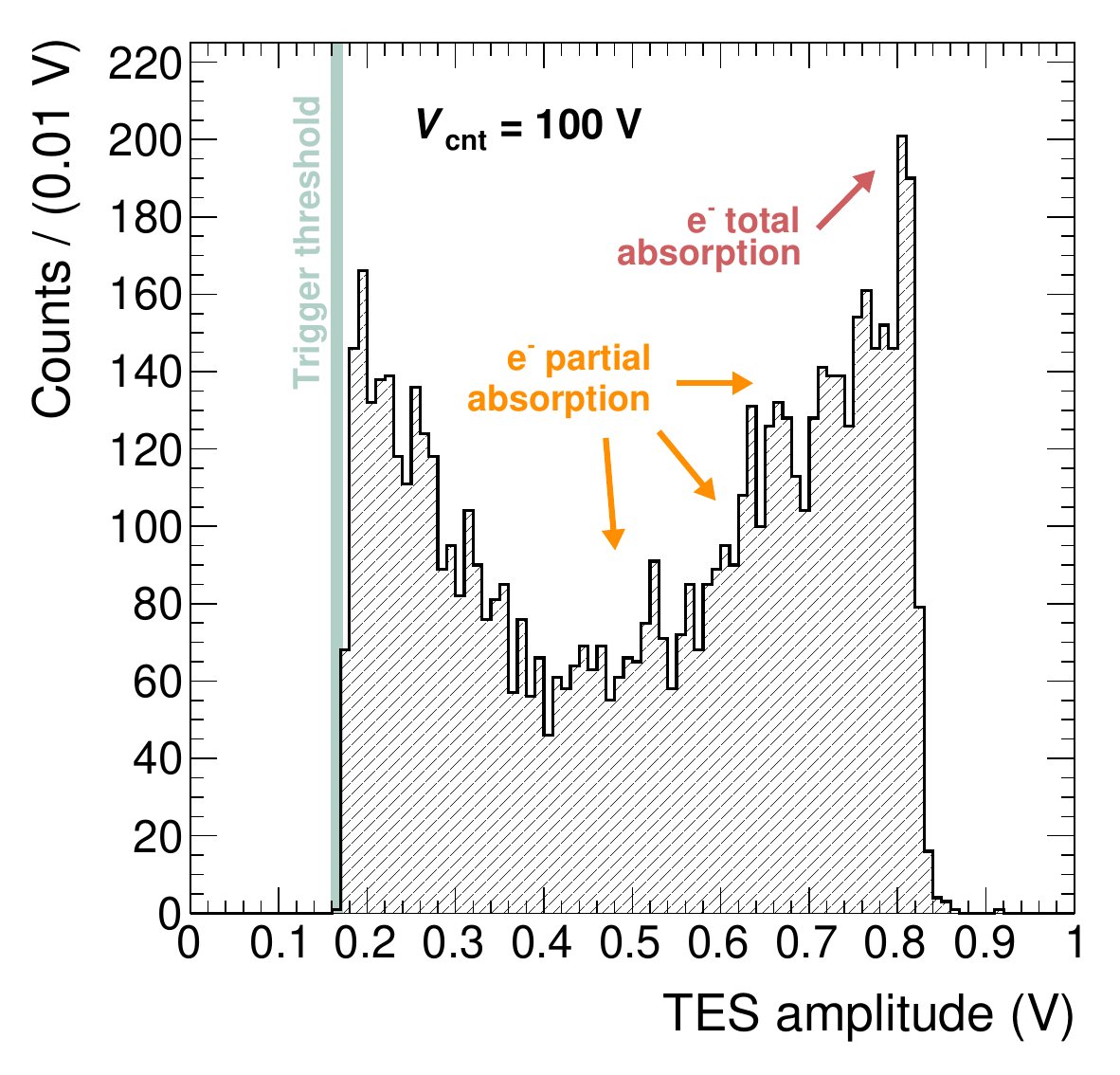}
\caption{Typical spectrum of TES signal amplitudes. This spectrum was obtained with $V_{\mathrm{cnt}} = 100$~V.\label{fig:spectrum}}
\end{figure}

For each value of $V_{\mathrm{cnt}}$, the amplitude of the signals produced by the TES were analyzed. A typical spectrum, obtained for $V_{\mathrm{cnt}} = 100$~V is shown in Figure~\ref{fig:spectrum}: as can be seen the distribution presents a high-amplitude peak, corresponding to the full absorption of the electron energy in the sensitive layer of the TES; a marked tail to the left of the peak, due to partial absorption of electrons, and is most likely due to electrons which fail to be stopped (`punch-through') by the thin TES bilayer (45~nm); and a low-amplitude peak, truncated by the trigger threshold of 166~mV, compatible with electrons back-scattered out of the TES after exciting an internal mode in the Au layer.

\begin{figure}[tb]
\includegraphics[width=0.4\textwidth]{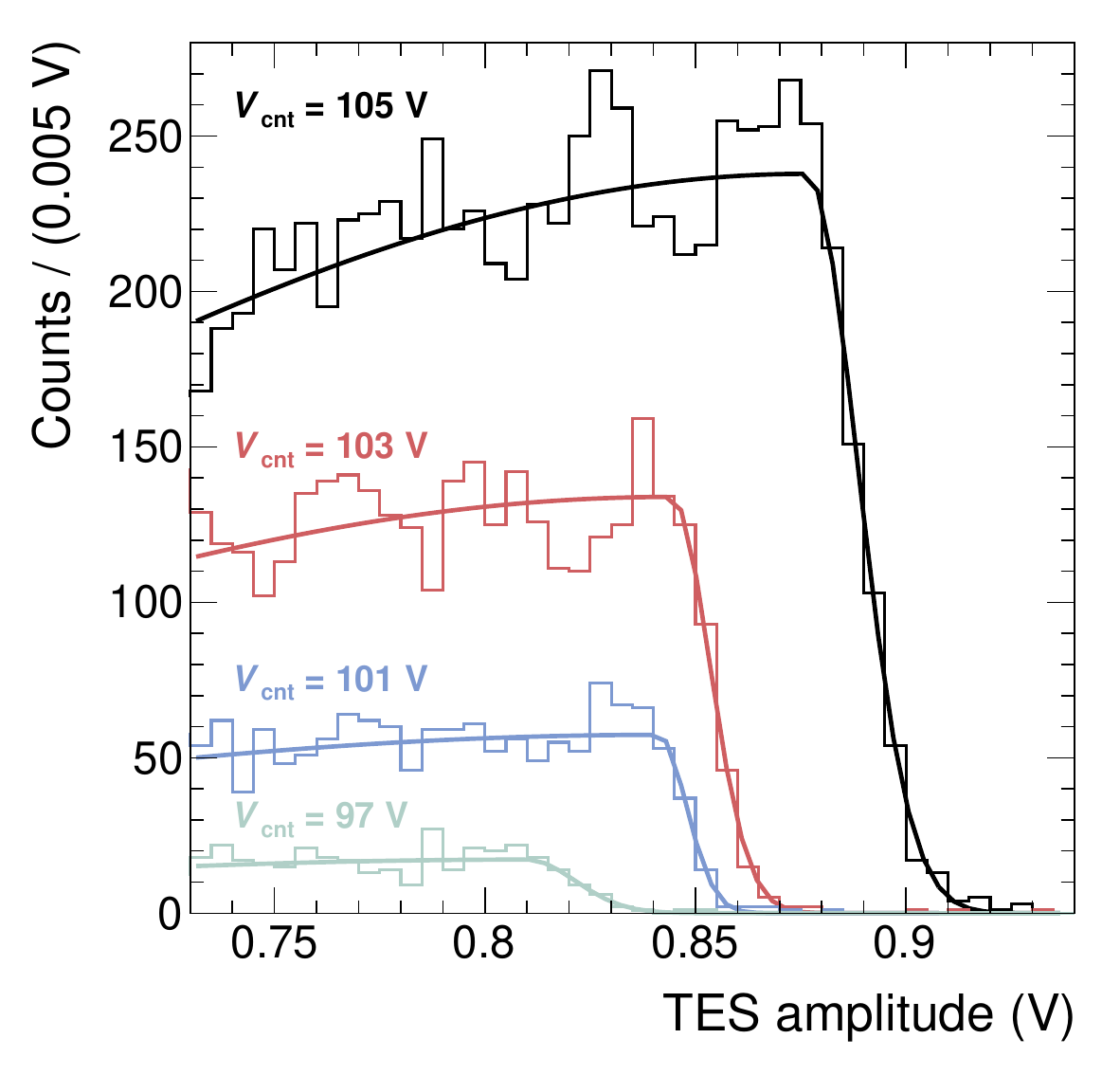}
\caption{Example fits, for four different values of $V_{\mathrm{cnt}}$, of the high-amplitude peak with the asymmetric Gaussian function.\label{fig:fits}}
\end{figure}

We fit the high-amplitude peaks of these distributions with an asymmetric Gaussian function, described by its peak position $\mu$ and its left~($\sigma_{\mathrm{L}}$) and right~($\sigma_{\mathrm{R}}$) tails. Some example fits, for $V_{\mathrm{cnt}} = 97$, 101, 103, and 105~V, are shown in Figure~\ref{fig:fits}. 


%

\begin{figure}[tb]
\includegraphics[width=0.4\textwidth]{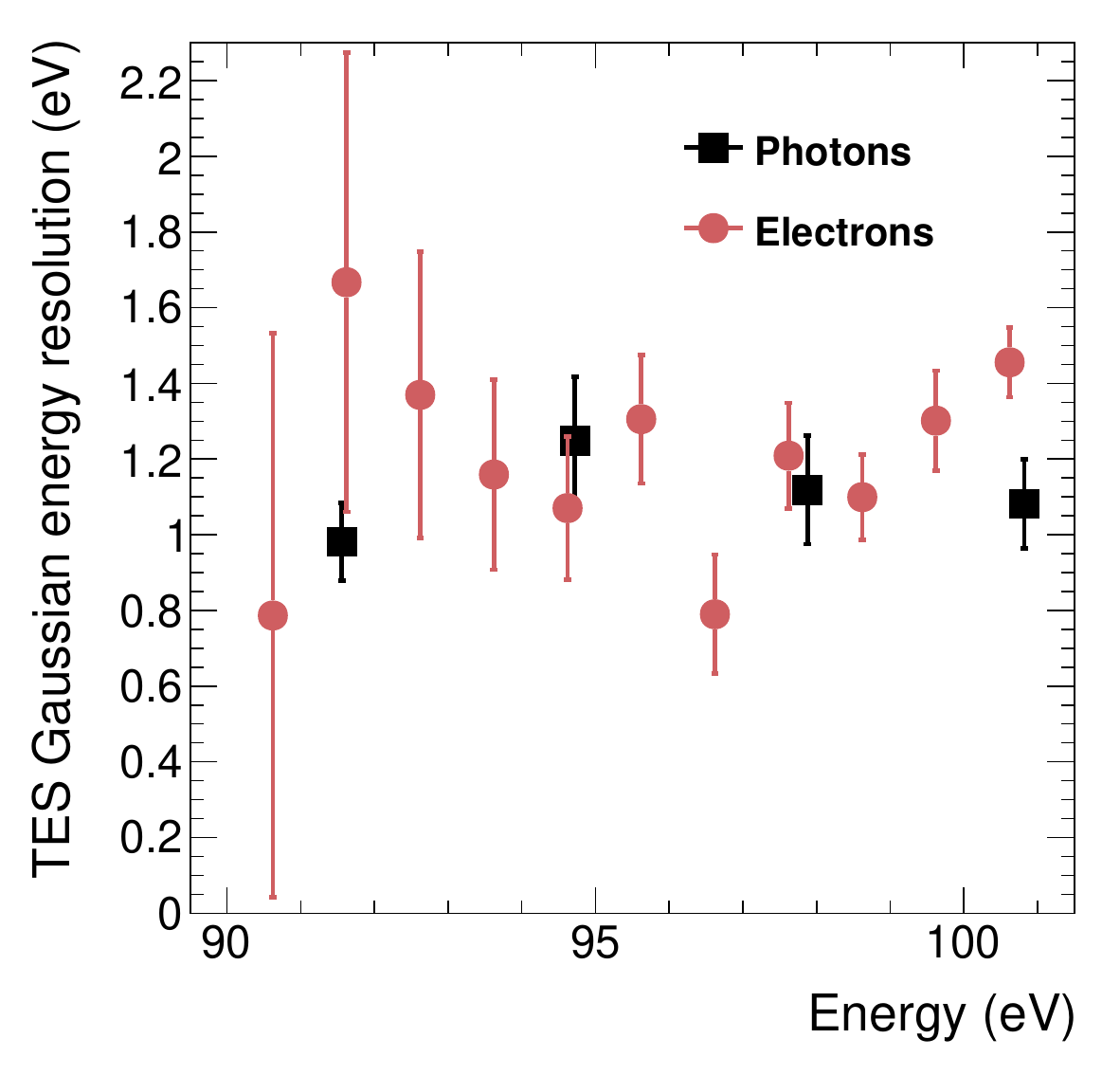}
\caption{
TES electron energy resolution for photons~(black square markers) and electrons~(red circular markers) as a function of the particle energy.\label{fig:mu}}
\end{figure}


Figure~\ref{fig:mu} shows the TES Gaussian energy resolution for electrons (red circular markers) and photons~(black square markers) as a function of the incoming particle energy. The photon energy resolution is defined as the Gaussian widths of the fit shown in Figure~\ref{fig:photons}. The electron energy resolution is defined as  $\sigma_{\mathrm{e}} = (\sigma_{\mathrm{R}}/\mu)\cdot E_{\mathrm{e}}$, where the electron kinetic energy $E_{\mathrm{e}}$ is defined in Equation~(\ref{eq:energy}). The parameter $\sigma_{\mathrm{R}}$ describes the broadness of the high-energy tail of the absorption peak in the TES amplitude distribution, which is  dominated by the energy resolution of the device, plus negligible effects due to the non-monochromaticity of the source. 


The electron Gaussian resolution is measured to be $0.8 < \sigma_{\mathrm{e}} < 1.8$~eV for electrons in the $90 \leq E_{\mathrm{e}} \leq 101$~eV energy range, and is found to be compatible with the energy resolution of photons in the same energy range. This is a non-trivial result, as at such low energy these particles have different interactions with matter, and suggests that the the heat-based detection mechanism in the TES is the same for electrons and photons. This would in turn cast optimism on the PTOLEMY target of $\sigma_{\mathrm{e}} = 50$~meV, because this target has already been achieved by these devices with photons \cite{lolli}. The focus of future work will therefore be on lowering the energy of the incoming electrons, and this can be done by decreasing the source-sensor distance $d$~(thus lowering the $V_{\mathrm{cnt}}$ at which FE starts), or by introducing decelerating potentials.

To conclude, this work reports the detection of low-energy electrons with a transition-edge sensor. This has been achieved with the use of an innovative cold electron source, based on field emission from vertically-aligned carbon nanotubes.  We obtain a Gaussian energy resolution between 0.8 and 1.8~eV for fully-absorbed electrons in the $(90-101)$~eV energy range, which matches the energy resolution of the same TES device for photons in the same energy range. This work has significant implications for $\beta$-decay and neutrino physics, where many experiments would benefit from high-resolution, low-energy electron detection.



\section[*]{Acknowledgements}
The authors are grateful to Elio Bertacco, Martina Marzano, Matteo Fretto and Ivan De Carlo for technical support, as well as the PTOLEMY collaboration for useful discussions. This research was partially funded by the contribution of grant 62313 from the John Templeton Foundation, by PRIN grant `ANDROMeDa' (PRIN\_2020Y2JMP5) of Ministero dell'Universit\`a e della Ricerca, and from the EC project ATTRACT (Grant Agreement No. 777222). 

\end{CJK*}

\bibliographystyle{apsrev4-1}
\bibliography{references}

\end{document}